\documentstyle[twoside,fleqn,espcrc2,epsf]{article}
\global\let\epsfloaded=Y

\newcommand{\AmS}{{\protect\the\textfont2
  A\kern-.1667em\lower.5ex\hbox{M}\kern-.125emS}}

\title{Possible Resonances in $\mu^+e^- \to \mu^- e^+$ Collisions}

\author{George W. S. Hou
 \address{Department of Physics, National Taiwan University, \\
                   Taipei, Taiwan 10764, R.O.C.}%
   \thanks{Work supported by grant NSC 85-2112-M-002-011 of the Republic of China,
           and in collaboration with G. G. Wong.}
}

\begin{document}

\begin{abstract}
We study the possibility of discovering resonances in 
$\mu^+e^- \to \mu^- e^+$ and $e^-e^- \to \mu^- \mu^-$ collisions.
We begin with the closely related problem
of  muonium--antimuonium transitions,
where the experimental limit has just been
improved by one order of magnitude.
We show that the new limit enters a rather interesting mass
and Yukawa coupling domain for neutral scalar bosons.
The stringent $\mu\to e\gamma$ decay is evaded by invoking some
multiplicative lepton number.
Neutral and doubly charged scalar bosons
give rise to distinguishable effects in
muonium transitions.
Alternatively, they could show up as spectacular  resonance peaks
in high energy $\mu^+e^- \to \mu^- e^+$and $e^-e^- \to \mu^- \mu^-$ collisions, 
respectively.
This could occur independent of whether, but especially when,
muonium-antimuonium transitions are experimentally observed.
\end{abstract}

\maketitle

\section{MOTIVATION }

Our title may appear to be somewhat exotic,
perhaps even a bit ``against gravity".
In fact we are not advocating that $\mu^+e^-$ resonances are likely,
but rather just urging for their consideration,
especially in the context of planning for muon colliders.
Let us, however, try to give {\it some} justification:

\subsection{Why Not?}

Just like the $\gamma\gamma$ and $e^- e^-$ collider options for
the NLC, as one considers the $\mu^+ \mu^-$ collider,
one should entertain the $\mu^+ e^-$
(or, $\mu^+\mu^+$, etc.) option as well.
Is it {\it in principle} not feasible? Or is it too costly?
Could it in fact be more easily achieved
(half a $\mu^+ \mu^-$ facility) hence cheaper?
I myself certainly cannot give any answer to this,
but perhaps some expert here could write things off from the outset \ldots

\subsection{Recent Progress in $M$--$\bar M$ Conversion}

This is a somewhat esoteric subject in itself, not-so-well-known,
but it consists of a series of beautiful low energy experiments which
use $\mu^+$ to form muonium ($M \equiv \mu^+ e^-$ atom).
Clearly there is a connection between $M$--$\bar M$
conversion and high energy $\mu^+e^- \to \mu^- e^+$ collisions.

On the experimental side, if one considers the {\it published} limit \cite{Matthias},
which dates from 1991, the limit on the effective 4-fermi coupling
is still rather poor,
\begin{equation}
G_{M\bar M} < 0.16\, G_F.
\end{equation}
This limit, however, has just been improved \cite{MMBAR} to $0.018\, G_F$
by the MMBAR Collaboration at PSI \cite{PSI},
and is expected to reach the $10^{-3}\, G_F$ level in the course of
two years  \cite{Jungmann}.
On the theory side, recent activities have lead to new models.

We shall thus consider what is known about
$M$--$\bar M$ conversion in some detail before we turn towards
the question of resonances in $\mu^+e^- \to \mu^- e^+$ collisions
at the end.

\section{HISTORIC INTERPLAY}

\subsection{Muonium--Antimuonium Conversion}

Following the suggestion of
$K^0$ -- $\bar K^0$ mixing by Gell-Mann and Pais,
 in 1957 Pontecorvo \cite{Pont} pointed out that
the $M$ -- $\bar M$  system, which are {\it atomic} states of
$\mu^+ e^-$ and $\mu^- e^+$, would be ideal for testing the
neutral particle-antiparticle mixing idea.
Unlike the hadronic case, the matrix elements are fully calculable.

Muonium was formed for the first time by Vernon Hughes
and his team in 1960.
In 1961, Feinberg and Weinberg \cite{FeinWein} wrote down,
in analogy
to the $V - A$ weak interactions,
the reference standard interaction of the form
\begin{equation}
{\cal H}_{M\bar M} = {G_{M\bar M} \over \sqrt{2}}
                                         \bar \mu \gamma_\lambda (1-\gamma_5) e \,
                                         \bar \mu \gamma^\lambda (1-\gamma_5) e + \mbox{h.c.}
\end{equation}
Experimental results have since been given \cite{PDG} in terms of
upper limits on $G_{M\bar M}$ in comparison with $G_F$.
It was also 
noted that although $M$--$\bar M$ conversion is forbidden
by the usual additive lepton number conservation,
it is allowed by the possibility 
of {\it multiplicative} muon number,
which is analogous to Pais' original suggestion for strangeness.
Shortly thereafter, Glashow \cite{Glashow} noted
that, complementary to $M$--$\bar M$ conversion,
$e^- e^- \to \mu^-\mu^-$ collisions
could also provide
interesting tests of multiplicative lepton number.

With $M$ formed in gas,
the first experiment on $M$--$\bar M$ conversion
lead to the limit \cite{Amato} of
\begin{equation}
G_{M\bar M} < 5800\, G_F \ \ \ \ \ \ \ \ \ \ \ \ \ \ \ \ \ \ \ (1968)
\end{equation}
This was improved by an order of magnitude within a year \cite{Barber}
by the pioneering $e^- e^-$ collider  experiment at SLAC
via  searching for $e^- e^- \to \mu^-\mu^-$.
Unfortunately, this track has never been repeated,
and further experiments were all along the $M$--$\bar M$ conversion line.

The 1968--1969 limits were so weak, they indicate
clearly both the experimental challenge and the
wide open possibilities!
But experiment lay dormant for 13 years,
in part because of ``interruptions" (e.g. at SLAC)
during the heady days  of 1969 -- 1975,
and in part because activities turned towards
$\mu\to e\gamma$ type of experiments at low energy facilities.
In 1982, the new limit of
$G_{M\bar M} < 42\, G_F$ was established by a TRIUMF experiment,
which utilized $M$ formation in vacuum.
Continued improvements went on for a decade,
employing methods such as thermal $\mu^+$ techniques,
or detecting the slow atomic $e^+$ from $\bar M$ after
$\mu^-$ decay, etc.
Finally, studies at  LAMPF lead to the limit of eq. (1).

The MMBAR Collaboration at PSI employs the same techniques as
the LAMPF experiment, but with a factor of $\sim 300$
improvement in acceptance,
and detecting annihilation photons by transporting the slow $e^+$
in a controlled way.
As such, it aims to both reduce the background and drastically
improve the reach. 
The new limit of 
\begin{equation}
G_{M\bar M} < 0.018\, G_F,
\end{equation}
demonstrates that it is feasible \cite{Jungmann}
for MMBAR to reach its goal of $10^{-3}\, G_F$ sensitivity,
provided it gets sufficient run time.
As we shall see, these limits would start to cut into
the parameter space of interest for various new models.

\subsection{$e^-e^- \to \mu^- \mu^-$ Studies?}

It should be emphasized that the suggestion of Glashow was tested just once at SLAC in 1969,
as a by-product of the QED test with $e^- e^-$ collisions.
Since then $e^\pm$ beams and collider technology
has improved dramatically,
but $e^-e^- \to \mu^- \mu^-$ search has never been repeated.
Theorists, however, have on and off made proposals
for the utility of such studies, as well shall see later.

\subsection{Additive vs. Multiplicative Law}

By now the additive lepton number conservation law,
by convention,has become part of the Standard Model.
However, since the additive law appears to emerge by accident,
it should not be held as more sacred than the multiplicative law.
The pursuit of experimental tests of such laws
is therefore of {\it fundamental} nature.
In Table 1 we make some comparison between additive vs. multiplicative law
for several relevant processes.
It is clear that the additive law is more restrictive than the multiplicative one,
and there are three places \cite{wrong_nu} to test and distinguish the two.
Of course, $M$--$\bar M$ conversion is a particular
low energy form of $\mu^+e^- \to \mu^- e^+$ transition.
Although one has a rather stringent limit on $\mu\to e\gamma$,
in a way, the 1977 rumors at PSI regarding this mode provided some
inspiration for resuming muonium conversion studies.
Alternatively, since 1982, renewed experimental interests
stimulated theoretical work, and
new interactions started to appear in a few categories.

\begin{table}[hbt]
\newlength{\digitwidth} \settowidth{\digitwidth}{\rm 0}
\catcode`?=\active \def?{\kern\digitwidth}
\caption{
Contrast between additive and multiplicative lepton number
for various exotic processes.}
\begin{tabular}{
lcc}
\hline
                &  Additive
                &  Multiplicative    \\
\hline
$\mu\to e\gamma$                &    \boldmath{\large $\times$}   &     \boldmath{\large $\times$}   \\
$\mu^+e^- \to \mu^- e^+$ 
 &    \boldmath{\large $\times$}   &     $\surd$   \\
$e^-e^- \to \mu^- \mu^-$      &    \boldmath{\large $\times$}   &     $\surd$   \\
$\mu^+ \to e^+ \bar\nu_e \nu_\mu$
                                                 &    \boldmath{\large $\times$}   &     $\surd$   \\
\hline
\end{tabular}
\end{table}

\section{RECENT MODELS}

Besides the work of Yoshimura and coworkers  \cite{Yoshimura}, 
as well as Derman and Jones \cite{Derman},
there was little theoretical activity after the paper of Feinberg and Weinberg.
However, stimulated by experimental and theoretical progress,
an important piece of work was advanced regarding
doubly charged Higgs bosons.

\subsection{$\Delta^{++}$: Doubly Charged Higgs}

The left-right symmetric model of 
Mohapatra and Senjanovic \cite{MS}
contained Higgs triplets that can be put in the form
\begin{equation}
\left[ \begin{array}{cc}
	\Delta^+/\sqrt{2} & \Delta^{++} \\
	\Delta^0 & -\Delta^+/\sqrt{2} 
	\end{array}
\right],
\end{equation}
which naturally violates lepton number.
In 1982 Halprin \cite{Halprin} pointed out
that the doubly charged scalar $\Delta^{++}$
can mediate $M$--$\bar M$ conversion {\it at tree level}
via the effective interaction
\begin{equation}
{\cal H}_{\Delta^{++}} = {f_{ee} f^*_{\mu\mu} \over 4 m^2_{\Delta^{++}}}\,
                                         \overline{e^c} (1-\gamma_5) e \,
                                         \bar \mu (1-\gamma_5) \mu + \mbox{h.c.}
\end{equation}
After Fierz rearrangement, this can be put into the $(V-A)(V-A)$ form
of eq. (2), where the coefficient in eq. (6)
is identified with $\sqrt{2}\, G_{M\bar M}$.
The importance of this work is that
it provided some {\it gauge} interaction foundation for the
$(V-A)(V-A)$ form of Feinberg and Weinberg.

Further experimental results stimulated an interesting twist
that ``unwound" the above ``gauge" motivation.
In 1989, Chang and Keung \cite {CK} proposed a generic model
for  $\Delta^{++}$, stripped of the LR symmetry motivation,
but resurrecting Feinberg-Weinberg's notion of
multiplicative lepton number to forbid $\mu \to e\gamma$.
Herczeg and Mohapatra \cite {HerMoh}, however,
continued along the traditional line.
By making some assumptions on $m_{\nu_\mu}$, etc., 
they argued for a lower bound of
$G_{M\bar M} > 7\times 10^{-5}\, G_F$ in the context of LRS model.

Here we see the classic situation of mutual stimulation between
experiment and theory.

\subsection{$X^{++}$: Dilepton}

From a different perspective, and originally unaware of its contact
with muonium conversion,
Adler, and Frampton and Lee \cite{dilepton} proposed models
where left-handed $\nu_e$, $e^-$ and $e^+$ formed
a triplet under a gauged
$SU_\ell(3)$ symmetry, which could arise from an underlying
$SU(15)$ symmetry.
The theory possessed lepton number violating $X^+$ and
$X^{++}$ gauge bosons called dileptons.
Frampton \cite{Frampton}, in particular, 
called for $e^-e^- \to \mu^- \mu^-$ collider studies. 
However, Sasaki and coworkers \cite{Fujii} subsequently showed
that the dilepton could mediate $M$--$\bar M$ conversion, 
but the effective Hamiltonian is of $(V-A)(V+A)$ form!
One now has a second gauge theory motivation for muonium conversion,
but in a form {\it different} from that of eq. (2).
It took some time for experimenters to appreciate its significance.

\subsection{$\tilde\nu_\tau$: Tau Sneutrino}

In the context of SUSY models with
two $R$-parity violating couplings,
Halprin and Masiero \cite{HM} suggested that
one has a unique $L$-parity violating coupling
of superfields in the leptonic sector of the form
$L_i L_j E_k^c$, and the tau sneutrino $\tilde\nu_\tau$
could mediate $M$--$\bar M$ conversion.
They find that 
\begin{equation}
{\cal H}_{\tilde\nu_\tau} =
                                 -{\lambda_{312} \lambda^*_{321} \over m^2_{\tilde\nu_\tau}}\,
                                         \overline{\mu} (1+\gamma_5) e \,
                                         \bar \mu (1-\gamma_5) e + \mbox{h.c.}
\end{equation}
Note that this is of $(S\pm P)(S\mp P)$ form, a point that was 
not pursued by the authors.
Note also that $\tilde\nu_\tau$ is a special case of
complex {\it neutral} scalars, 
rather than doubly charged bosons of the previous two cases.
In some sense, this may be less obscure
since neutral scalars already appear in the Standard Model.

\subsection{$\Phi^0 = H$ or $A$: Neutral Scalar}

Motivated by the mass hierarchy problem,
we \cite{WH} pursued the possibility where
lower generation charged lepton masses were generated
radiatively by simple one loop diagrams.
As illustrated in Fig. 1, this involves flavor changing
neutral scalar bosons
$H$ and $A$
\begin{figure}[htb]
\let\picnaturalsize=N
\def\picsize{2.0in}
\def\picfilename{mumu95_1.eps}
\ifx\nopictures Y\else{\ifx\epsfloaded Y\else\input epsf \fi
\global\let\epsfloaded=Y
\vskip -.4cm
\centerline{\ifx\picnaturalsize N\epsfxsize \picsize\fi \epsfbox{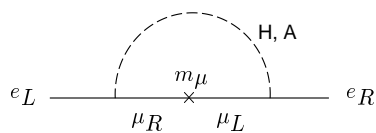}}}\fi
%
\vskip -1.4cm
\caption{Radiative mass generation mechanism.}
\end{figure}
\begin{figure*}[hbt]
%
\let\picnaturalsize=N
\def\picsize{4.0in}
\def\picfilename{mumu95_2.eps}
\ifx\nopictures Y\else{\ifx\epsfloaded Y\else\input epsf \fi
\global\let\epsfloaded=Y
\centerline{\ifx\picnaturalsize N\epsfxsize \picsize\fi \epsfbox{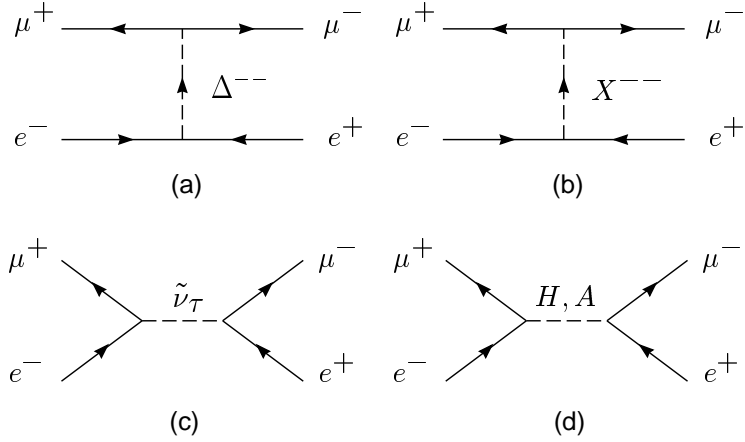}}}\fi
%
\vskip -0.7cm
\caption{Mechanisms for inducing $\mu^+e^- \to \mu^- e^+$ transitions
at tree level, as discussed in the text.}
\end{figure*}
(scalar and pseudoscalar, respectively),
with the next generation lepton as ``seed" mass.
It was crucial to have
non-degeneracy of the two scalars, hence this is a case
of ``broken" complex scalar.
$H$ and $A$ mediate
$SS$ and $PP$ effective interactions, respectively,
which are linear superposition of 
$(S\pm P)(S\mp P)$ and $(S\pm P)(S\pm P)$,
hence the previous $\tilde\nu_\tau$ mediated effect is indeed
a very special case.
We shall discuss the simplified model 
in more detail in the following section.

We summarize in Fig. 2 the four types of interactions that
could mediate $M$--$\bar M$ conversion at {\it tree level}.
The doubly charged bosons are exchanged in $t$-channel,
while both $s$- and $t$-channels (not shown) are possible for
neutral bosons.
Note that ``horizontal" neutral gauge bosons 
(as originally mentioned by Feinberg and Weinberg)
may also be possible,
but seem difficult to construct.

We turn to a more detailed account of the neutral scalar case,
as it is most interesting and relevant for $\mu^+e^-$ collisions.
We shall concentrate on $M$--$\bar M$ conversion 
in the following section.

\section{CASE STUDY: $\Phi^0 = H$ OR $A$}

\subsection{Operators and Matrix Elements}

Suppose
neutral scalar and pseudoscalar bosons $H$ and $A$ exists
and have the couplings,
\begin{equation}
-{\cal L}_{Y} = {f_H \over \sqrt{2}}\, \bar \mu e\, H
                     + i {f_A \over \sqrt{2}}\, \bar \mu \gamma_5 e \, A + H.c.
\end{equation}
Introduce \cite{CK} electron parity $P_e$,
such that the fields $e$, $H$, $A$ $\longrightarrow -e$, $-H$, $-A$
hence odd under $P_e$, while $\mu$ and all other fields are even.
Then processes odd in number of electrons (plus positrons)
like $\mu\to e\gamma$ and $\mu\to ee\bar e$ are forbidden.
$P_e$ is nothing but a variation of
the multiplicative muon parity of
Feinberg and Weinberg \cite{FeinWein}.
Note that $H$ and $A$ are rather exotic in that they
possess FCNC couplings {\it only}.
The interaction of eq. (8) leads to the low energy effective Hamiltonian
\begin{equation}
{\cal H}_{S,P} = {f_H^2 \over 2m_H^2}\, \bar \mu e\, \bar \mu e
                           - {f_A^2 \over 2m_A^2}\, \bar \mu \gamma_5 e\, \bar \mu \gamma_5 e,
\end{equation}
which is relevant for mediating $M$--$\bar M$ conversion.
This can be rewritten as
\begin{eqnarray}
{\cal H}_{S,P} = \left({f_H^2 \over 4m_H^2} + {f_A^2 \over 4m_A^2}\right)
                                                                                     (S-P)(S+P) \ \ \ \ && \nonumber \\
           + \left({f_H^2 \over 4m_H^2}  - {f_A^2 \over 4m_A^2}\right)
                                                \left[{1\over 2}\, (S-P)(S-P) \right. \ \,  && \nonumber  \\
                                                \left. + {1\over 2}\, (S+P)(S+P) \right]. &&
\end{eqnarray}
The first term is analogous to the $\tilde\nu_\tau$ induced effect
discussed above,
while the second term is subdominant because of a suppressed coefficient.
Three special cases can be considered:
\begin{enumerate}
\item Pure scalar: $f_A = 0$, which is of $SS$ form.
\item Complex scalar: $f_A = f_H$, $m_A = m_H$,
	hence of $SS - PP$ form, as in the $\tilde\nu_\tau$ case.
\item Pure pseudoscalar: $f_H = 0$, hence $PP$ form.
\end{enumerate}

The $M \longrightarrow \bar M$ transition matrix element
can be written as,
\begin{eqnarray}
{\cal M}_{s_{\bar e}\,s_\mu; \; s_e\,s_{\bar \mu}} &=& 
                  \langle \bar M_{s_{\bar e}\,s_\mu} \vert {\cal H}_{M\bar M}
                                                            \vert M_{s_e\,s_{\bar \mu}} \rangle
                                                                    \equiv \delta/2  \nonumber \\
    &=& C\, \vert \Psi_{1S}(0)\vert^2\; T_{s_{\bar e}\,s_\mu; \; s_e\,s_{\bar \mu}}  
\end{eqnarray}
where, for simplicity, we take $\delta$ to be real \cite{FeinWein}.
The notation in eq. (11) is more or less self-explanatory:
$C$ is the effective coupling, $\Psi(0)$ is the wave function at the origen,
while the $T$ matrix describes spin mapping.
In the nonrelativistic limit,
with $\bf p$ $ \to 0$, one is left with only spin degree of freedom.
For example, for $SS$ and $PP$ cases one finds
\begin{eqnarray}
T^{SS}_{s_{\bar e}\,s_\mu; \; s_e\,s_{\bar \mu}}
    &=& - \;\bar v(s_{\bar \mu}) v(s_{\bar e}) \bar u(s_\mu) u(s_e)  \nonumber \\
    &  &+ \;\bar u(s_\mu) v(s_{\bar e}) \bar v(s_{\bar \mu}) u(s_e) \nonumber \\
    &  &+ \;\bar v(s_{\bar \mu}) u(s_e) \bar u(s_\mu) v(s_{\bar e}) \nonumber \\
    &  & - \;\bar u(s_\mu) u(s_e) \bar v(s_{\bar \mu}) v(s_{\bar e}) \nonumber \\
    &=& -\;2\, \delta_{s_\mu s_e} \delta_{s_{\bar \mu} s_{\bar e}}, 
\end{eqnarray}
\begin{equation}
T^{PP}
    = -2\, (\delta_{s_\mu s_{\bar \mu}} \delta_{s_{\bar e} s_e}
               - \delta_{s_\mu s_e} \delta_{s_{\bar \mu} s_{\bar e}}),
\end{equation}
whereas the usual $(V-A)(V-A)$ result is
\begin{equation}
T^{(V-A)(V-A)}
    = 8\, \delta_{s_\mu s_{\bar \mu}} \delta_{s_{\bar e} s_e}.
\end{equation}
Likewise, one easily finds 
\begin{equation}
T^{(V-A)(V+A)}
    = -4\, (\delta_{s_\mu s_{\bar \mu}} \delta_{s_{\bar e} s_e}
               - \delta_{s_\mu s_e} \delta_{s_{\bar \mu} s_{\bar e}}),
\end{equation}
\begin{equation}
T^{S^2-P^2}
    = 2\, (\delta_{s_\mu s_{\bar \mu}} \delta_{s_{\bar e} s_e}
               - \delta_{s_\mu s_e} \delta_{s_{\bar \mu} s_{\bar e}}),
\end{equation}
\begin{equation}
T^{S^2+P^2}
    = -2\, (\delta_{s_\mu s_{\bar \mu}} \delta_{s_{\bar e} s_e}
               - \delta_{s_\mu s_e} \delta_{s_{\bar \mu} s_{\bar e}}).
\end{equation}
Note that there are only two types of spin matching,
namely, $s_{\bar e}= s_e$, $s_\mu=s_{\bar \mu}$, or
$s_{\bar e}=s_{\bar \mu}$, $s_\mu= s_e$.
This is not surprising because of nonrelativistic reduction.
Note further that 
$T^{S^2+P^2} \propto T^{(V-A)(V-A)}$
and $T^{S^2-P^2} \propto T^{(V-A)(V+A)}$.
The former comes as a special consequence of the NR limit,
while the latter is because the two operators are Fierz related.

As shown by Feinberg and Weinberg \cite{FeinWein},
the time integrated probability for an initial state $M$
to decay as $\bar M$ is
\begin{equation}
P(\bar M) = {\delta^2/2 \over \lambda^2 + \Delta^2 + \delta^2},
\end{equation}
where $\lambda$ is the $\mu$ decay width,
and $\Delta = E_M - E_{\bar M}$ is the splitting between
$M$ and $\bar M$ energy levels for the given spin configuration.
Eq. (18), of course, is rather similar to the formula for
$K^0$--$\bar K^0$ mixing,
and one could have called muonium conversion
$M$--$\bar M$ oscillations.
What is quite different from the $K^0$ -- $\bar K^0$ system
is that $P(\bar M)$ has strong $B$ field dependence \ldots

\subsection{Magnetic Field Dependence!}

The point is that $M$ and $\bar M$ are atomic states
that are much larger in size than hadrons,
hence $E_M -  E_{\bar M}$ is sensitive to the material environment!
For example, in gas, solids etc. 
strong local fields lead to large splittings \cite{FeinWein},
that is why one has to form and study muonium in vacuum.

Dependence on ambient $B$ field enters through
the well known Zeeman effect, which leads to 
mixing and level repulsion between different spin configurations.
The $B$ field dependence of $P(\bar M)$ for the $(V-A)(V-A)$ case
was known to the experimenters, since it was
calulated by Feinberg and Weinberg.
In reporting experimental limits on $G_{M\bar M}$ one in fact 
routinely corrects for this,
since $B$ fields are present in the apparatus
for sake of charged particle tracking.
However, since the $(V-A)(V-A)$ form has long been the
reference standard in this business,
and in part because the work of Halprin and others gave
the phenomenological $(V-A)(V-A)$ interaction some gauge foundation,
the $B$ dependence of the transition probability became
a routine technical adjustment for the experimenters,
while theorist generally were unaware of it.
Therefore when the
dilepton induced effective interaction
was found to be of $(V-A)(V+A)$ form,
out of inertia one had a tendency to make the same 
correction.

It was quite interesting, therefore, 
when Horikawa and Sasaki pointed out \cite{Sasaki}
that the $B$ field dependence for
$(V-A)(V+A)$ interaction was different from $(V-A)(V-A)$ case.
In particular, 
it is not as suppressed by the typical 1kG field,
the field strength used in the MMBAR experiment.
This implies that limits on the effective coupling of $(V-A)(V+A)$
interactions are more stringent than the $(V-A)(V-A)$ case.

Stimulated by this work, we began a systematic study \cite{HW}
of $B$ field dependence for all possible types of interactions.
The formalism of the previous subsection can be readily used,
once one finds a proper way to incorporate the interaction of
electron and muon spin (moments) with the external $B$ field.
Note that  in the strong $B$ field limit, 
{\boldmath $\mu$}$_e$ and {\boldmath $\mu$}$_{\bar e}$
moments are aligned/anti-aligned with the {\boldmath $B$} field.
Clearly, for nonzero $B$ field, one should be using the 
Breit--Rabi basis for energy eigenstates.

Skipping the formalism, 
the essence of  proper treatment is as follows \cite{HW}.
In the presence of nonvanishing {\boldmath $B$} field, 
eq. (11) becomes
\begin{equation}
\langle \bar M \vert {\cal H}_{M\bar M} \vert M \rangle
        \ \mbox{\rule[-.11in]{0.1mm}{8mm}\,\,}_{\mbox{\boldmath $B$} \neq 0}
                            \equiv \bar\delta/2,
\end{equation}
with the notion that
\begin{enumerate}
\item $\vert M \rangle$ is a mixture of $\vert s_e,\ s_{\bar \mu} \rangle$ states,
	and likewise for $\vert \bar M \rangle$. Hence, $\bar \delta \neq \delta$.
	It is traditional to label the Breit-Rabi energy levels 
	with the weak field basis $\vert M;F,m_F \rangle$,
	i.e. the corresponding zero field hyperfine states. 
\item One has to keep close watch of level splitting $\Delta$ caused by 
	{\boldmath $B$}$\neq 0$.
\end{enumerate}
With $\delta$ replaced by $\bar \delta$ in eq. (18),
the $B$ field dependence of {\it both} $\Delta$ and $\bar \delta$
for different ``hyperfine" $(F,\; m_F)$ states determine the
$B$ dependence of $P(\bar M)$.
This redefinition of $\bar \delta$ simplifies things
enormously.
For example, as {\boldmath $B$}$\rightarrow \infty$,
$\bar \delta \rightarrow 0$ for the $(V-A)(V-A)$ case,
but does not vanish for the $(V-A)(V+A)$ case.
It may appear to be a bit counter-intuitive ,
but it reveals that {\it detailed study at different 
$B$ fields can 
distinguish different interactions from one another},
assuming that one has a signal.
Finally, one has to sum over different $(F,\; m_F)$
contributions weighted by their population
$\vert c_{F,m_F}\vert^2$, and arrive at
the total transition probability
\begin{equation}
P_T(\bar M) =  \sum_{F,m_F}\vert c_{F,m_F}\vert^2\, P^{(F,m_F)}(\bar M).
\end{equation}

\begin{figure*}
\let\picnaturalsize=N
\def\picsize{4.5in}
\def\picfilename{mumu95_3.eps}
\ifx\nopictures Y\else{\ifx\epsfloaded Y\else\input epsf \fi
\global\let\epsfloaded=Y
\vskip 0.2cm
\centerline{\ifx\picnaturalsize N\epsfxsize \picsize\fi \epsfbox{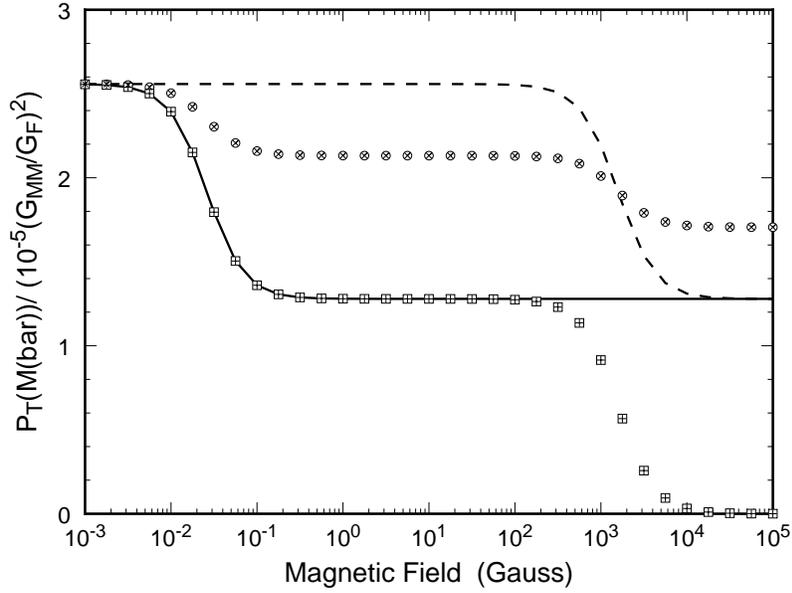}}}\fi
%
\vskip -2.3cm
\caption{
	Magnetic field dependence of total muonium conversion
             probability $P_T(\bar M)$ assuming 
              $\vert c_{1,  1}\vert^2 
             = \vert c_{1,  0}\vert^2
             = \vert c_{1,-1}\vert^2 
             = \vert c_{0,  0}\vert^2 
             = 1/4$, and
              normalized to conversion strength of $(V-A)(V-A)$ interaction
              at zero magnetic field.
              Solid and dashed lines stand for $SS$ and $PP$ operators,
              respectively,
              while $\circ$, {$\Box$}, {$+$} and 
              {$\times$}
              stand for $(S-P)(S+P)$, $(S-P)(S-P)$, 
              $(V-A)(V+A)$ and $(V-A)(V-A)$ cases, respectively.
}
\end{figure*}
\begin{figure*}
\let\picnaturalsize=N
\def\picsize{4.5in}
\def\picfilename{mumu95_4.eps}
\ifx\nopictures Y\else{\ifx\epsfloaded Y\else\input epsf \fi
\global\let\epsfloaded=Y
\vskip 0.3cm
\centerline{\ifx\picnaturalsize N\epsfxsize \picsize\fi \epsfbox{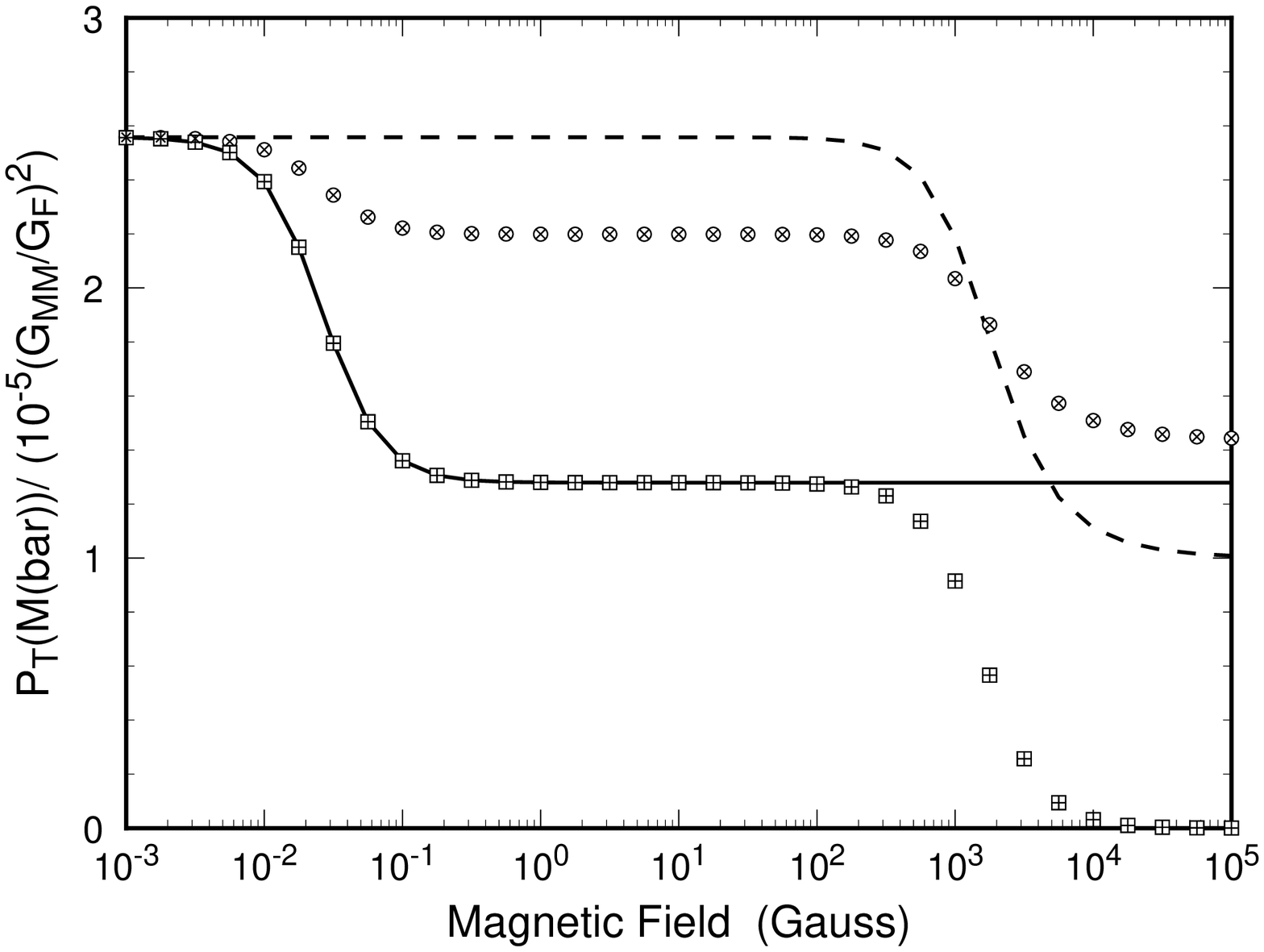}}}\fi
%
\vskip -2.3cm
\caption{
             Same as Fig. 3 except
              $\vert c_{1, 1}\vert^2 = 0.35$, 
              $\vert c_{1, 0}\vert^2 = 0.18$,
              $\vert c_{1,-1}\vert^2 = 0.15$ and
              $\vert c_{00}\vert^2 = 0.32$.
\vskip -.2cm
}
\end{figure*}

Without further ado, let us give the magnetic field dependence of
various muonium conversion interactions in Fig. 3,
assuming equally populated energy levels,
with notation as described in the long caption.
Note that the $(V-A)(V-A)$ (Feinberg and Weinberg)
and $(S-P)(S-P)$ cases behave identically,
and likewise for $(V-A)(V+A)$ (Horikawa and Sasaki)
and $(S-P)(S+P)$,
This can be understood from the proportionality among the
$T$ matrices in eqs. (14-17) as discussed previously.
The first dip starting at around a milli-Gauss
is due to the ``quenching" of $(1,\pm 1)$ modes
through the Zeeman splitting
(energy mismatch between $M$ and $\bar M$).
The purely pseudoscalar has vanishing matrix element in this
channel hence goes unsuppressed.
The second stage quenching effect come from
damping in the $(1,\, 0)$ and $(0,\, 0)$ modes,
where all interactions except pure scalar suffers some suppression,
with the $(S-P)(S-P)$ (and hence $(S+P)(S+P)$)
and $(V-A)(V-A)$ interactions drastically suppressed.
Thus, all other interaction types receive more stringent
bounds on their effective couplings for
experiments done with $B$ field $\sim$ 1kG.

We illustrate in Fig. 4 the same figure  but assuming that
$(0,\, 0)$, $(1, +1)$, $(1,\, 0)$, $(1, -1)$ are populated as
32\%, 35\%, 18\% and 15\%, respectively, \cite{Jungmann},
which supposedly corresponds to measured values
at 1.6kG magnetic field.
The results are not drastically different.

We refer to ref. \cite{HW} for further details.

\subsection{Present Constraints}

The published limit of Matthias et al. \cite{Matthias}, eq. (1),
is for $(V-A)(V-A)$ interaction at 10 Gauss.
The new limit of eq. (4) is also for $(V-A)(V-A)$ but at 1.6 kGauss.
One can read off from Figs. 3 or 4 the adjustments that one needs to
make in converting the bounds of eq. (1) and (4) to
other interaction cases.
For neutral scalars, we find
\begin{equation}
{f\over m} \ \raisebox{-.5ex}{\rlap{$\sim$}} \raisebox{.4ex}{$<$}\
            {1\over\ 1 \mbox{\rm TeV}},  {1\over\ 1.3 \mbox{\rm TeV}},
                               {1\over\ 1.6 \mbox{\rm TeV}},
\end{equation}
respectively, for $SS$, $PP$ and $(S-P)(S+P)$ interactions.
The latter occurs for the complex scalar (e.g. $\tilde \nu_\tau$) case.
Taking the Yukawa coupling $f\sim 1$, one can simply read off eq. (21)
to deduce mass bounds.
With smaller $f$, the mass bounds become weaker, hence more interesting.
Similar statements can be made for the doubly charged bosons.

Some other constraints on ${\cal H}_{S,P}$,
such as $(g-2)$ of electron and muon, 
or $e^+e^-\to \mu^+\mu^-$ compositeness search,
should be considered.
They turn out  \cite{HW2} to be all weaker than the MMBAR bound of
eq. (4), although the compositeness search constraint comes
surprisingly close.

\subsection{Implications of $M$--$\bar M$ Constraint}

It should be emphasized that the MMBAR constraint of
eq. (4) already  has some interesting implications on some models.
As mentioned earlier, our original motivation for venturing into
muonium conversion was the study of radiative mass generation
via flavor changing neutral scalars.
In the model of ref. \cite{WH},
scalar interactions of the type of eq. (8)
were used to generate charged lepton masses
iteratively order by order,
via effective one loop diagrams with
lepton seed masses from one generation higher.
A model was constructed where neutral scalars possessed
{\it weak scale} masses, 
while the Yukawa couplings $f\sim 1$ naturally

We are not concerned with
the generation of $m_\mu$ from $m_\tau$ here.
However, in analogy to the softly broken $Z_8$ symmetry
of ref. \cite{WH},
some discrete symmetry can be invoked to
forbid electron mass at tree level
but allow it to be generated by $m_\mu$
via one loop diagrams as shown in Fig. 1.
Since $m_{H,A} \gg m_\mu$, we have
\begin{equation}
   {m_e\over m_\mu}  \cong  {f^2\over 32\pi^2} \log{m_H^2\over m_A^2}.
\end{equation}
Note that $f_H = f_A = f$ is necessary for divergence cancellation,
hence in the $U(1)$ limit \cite{WH} of $m_A = m_H$
the mass generation mechanism is ineffective.
We see that, because the factor of $1/32\pi^2 \sim 1/300$
is already of order $m_e/m_\mu$,
if $m_A \neq m_H$ but are of similar order of magnitude,
in general we would have $f \sim 1$.
This looks attractive for scalar masses
far above the weak scale since one could
have large Yukawa couplings but
at the same time evade the bound of eq. (21).
However,
in the more ambitious model of ref. \cite{WH},
radiative mass generation mechanism is pinned to the weak scale,
namely, Higgs boson masses cannot be far above TeV scale
for sake of naturalness.
In this case, although eq. (22) still looks attractive and is a
simplified version of the more detailed results of ref. \cite{WH},
with $f \sim 1$ and $m_H,\ m_A <$ TeV,
the bound of eq. (21) cannot be satisfied.
We thus conclude that the new bound on $M$--$\bar M$ conversion \cite{PSI}
from PSI rules out the possibility
of radiatively generating $m_e$ {\it solely} from $m_\mu$
via one loop diagrams involving lepton number changing
neutral scalar bosons that have weak scale mass.
One would either need heavier scalars,
heavier seed masses (e.g. from $m_\tau$),
or $m_e$ has non-radiative origins.

\section{MUON-ELECTRON COLLIDERS?}

Recall that the $e^-e^- \to \mu^- \mu^-$ search suggested
by Glashow (and later by Frampton) has never been repeated
since 1969.
The $e^-e^-$ collider option is now being discussed under the
context of the NLC.
Perhaps we should start the discussion for $\mu^+e^-$ option
as we consider the $\mu^+\mu^-$ collider.

\begin{figure*}
\let\picnaturalsize=N
\def\picsize{4.5in}
\def\picfilename{mumu95_5.eps}
\ifx\nopictures Y\else{\ifx\epsfloaded Y\else\input epsf \fi
\global\let\epsfloaded=Y
\vskip 0.4cm
\centerline{\ifx\picnaturalsize N\epsfxsize \picsize\fi \epsfbox{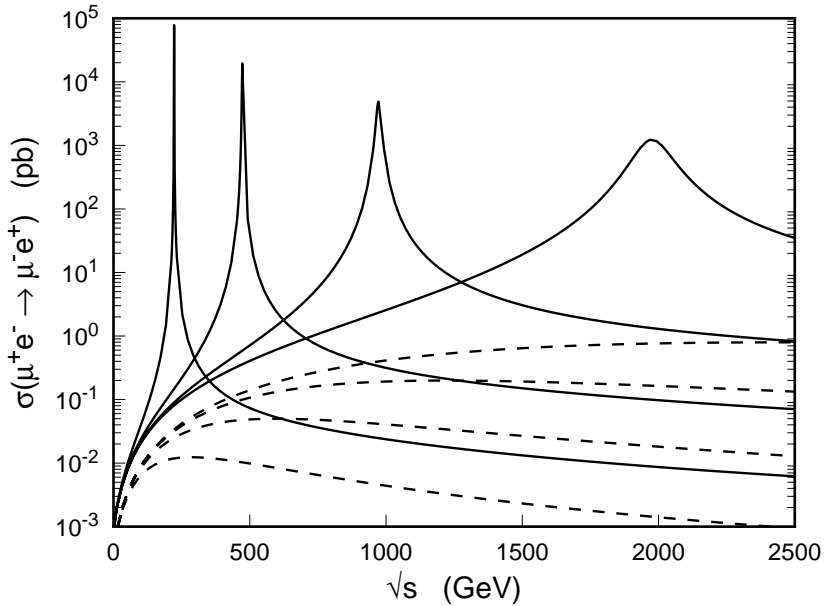}}}\fi
%
\vskip -2.3cm
\caption{$\sigma (\mu^+ e^- \rightarrow \mu^- e^+)\ \mbox{vs.} \ E_{\rm CM}$
         for $m_H =$ 0.25, 0.5, 1, 2 TeV.
        Only $H\to \mu^\pm e^\mp$ is taken into account for the width of $H$,
        with Yukawa couplings saturating
        eq. (21).
        Analogous bounds for the case of $\Delta^{--}$ are shown as dashed lines.
\vskip -.4cm
}
\end{figure*}

The low energy limits from MMBAR collaboration
imply that $f/m <$ 1/TeV, which is a rather interesting number.
It is plausible to have $f \sim 0.1 - 2$.
If such is the case, then scalars such as $H$, $A$ or $\tilde \nu_\tau$
could have mass $m > $100 GeV -- 2 TeV.
Not only this is not precluded by muonium conversion experiments, but
further results from these experiments would only improve
this mass bound by square root power.
If such bosons exist they would certainly show up as resonance peaks
(see Fig. 5) ) in $\mu^+ e^-$ collisons.
I am certainly very, very na\"\i ve,
but take, for example,
$E_e \sim 90$ GeV (LEP II energy),
and $E_{\mu^+} \sim$ 200 GeV -- 7 TeV (the latter is LHC energy),
then $\sqrt{s} \sim $ 200 GeV -- 1.1 TeV.
The signal is absolutely clean: $\mu^- + e^+$ production
from incoming $\mu^+ e^-$,
with $E(\mu^- e^+) = \sqrt{s}$.
Not only one has a large cross section, but there is 
practically zero background.

Now, if one sees a signal for muonium conversion
in the near future, one certainly would have to investigate further,
both by the means of $B$ field dependence of conversion rates,
but also working towards direct production of the bosons
responsible.
However, high energy search is rather complementary to the low energy one, for
as discussed above, even if the muonium conversion
limit improve by another order of magnitude and does not find
any signal, resonances in $\mu e$ channel are far from
ruled out.
As illustrated by the various (but somewhat scarce)
models, they could be related to mass generation or
to $R$-parity violation in SUSY.
Note that the doubly charged bosons
$\Delta^{++}$ and $X^{++}$ also give rise to non-negligible
cross sections.
These, of course, are more ideally searched for at $e^-e^-$
(or perhaps $\mu^+\mu^+$?) colliders.

Can such colliders, with a single $\mu^+$ beam
on an $e^-$ beam be done? Clearly some place like CERN
or SLAC (and NLC?) with existing high energy electron beams
would be at an advantage,
although I have the impression that one needs some
hadron facility to produce muons.
Perhaps performing $\mu^+ e^-$ collisions can be viewed as an end in itself.
Afterall, we should strive to collide together all kinds
of fundamental constituents of Nature.

\section{CONCLUSION}

We have {\it not} given, by any count,
compelling reasons that one should discover
resonances in $\mu^+e^-$ channel.
What we have presented is basically a review of the type of interactions
that could lead to lepton number violation in a different way than
the usual additive rule.
The results can be summarized in Table 2.
The doubly charged scalar emerges in some LRS models
with Higgs triplets.
The dilepton from some $SU_\ell(3)$ gauge theory
that puts $e^-$ and $e^+$ in the same lepton triplet.
Neutral scalars or pseudoscalars could
be related to flavor and the origin of mass,
while the tau sneutrino $\tilde \nu_\tau$ is
a special case of complex neutral scalar,
and could arise from SUSY models with $R$-parity violation.
Possibly except the doubly charged scalar, 
all other particles would be viewed by most people 
as rather exotic.
On the other hand, 
neutral scalars may be appealing since they already
appear in the standard model.
If any of these exist,
they could certainly show up as resonaces in
$\ell^\pm \ell^\pm$ ($\Delta^{++}$ or $X^{++}$),
or $\mu^+ e^-$ collisions ($H$, $A$ or $\tilde\nu_\tau$).
\vskip -0.4cm
\begin{table}[hbt]
\caption{
Agents for mediating $\mu^+e^- \to \mu^- e^+$ transitions.}
\begin{tabular}{
lcc}
\hline
  Spin$\backslash$Charge   &  $\vert Q\vert = 2$  &  $Q = 0$    \\
\hline
\ \ 1          &    $X^{++}$    &     Horizontal?   \\
\ \ 0          &    $\Delta^{++}$   &    $H$, $A$; $\tilde \nu_\tau$   \\
\hline
\end{tabular}
\end{table}
\vskip0.2cm

It should be emphasized that $\mu^+ e^- \to \mu^- e^+$
and $e^- e^- \to \mu^-\mu^-$ studies are quite complementary to
$M$--$\bar M$ conversion experiments.
If the latter discovers a signal in the near future,
then resonance studies in $\mu^+ e^- \to \mu^- e^+$
and $e^- e^- \to \mu^-\mu^-$ channels would be absolutely necessary.
However, if muonium conversion studies yield a negative result,
it does not imply that there cannot be lepton number violation
allowed by multiplicative rule occuring at the TeV scale.

The question for this meeting is, therefore,
whether there are any fundamental difficulties in colliding
$\mu^+$ on $e^-$.
Can such studies be an end in itself?
Afterall, we should collide together all possible fundamental
constituents of Nature in an effort to reveal its secrets.


\begin{thebibliography}{99}
\bibitem{wrong_nu}
	The $\mu^+ \to e^+ \bar\nu_e \nu_\mu$ process,
	so-called ``wrong neutrino" decay,
	has also been searched for experimentally (see ref. \cite{PDG}).
\end{thebibliography}
\end{document}